\documentclass[10pt,conference]{IEEEtran}
\usepackage{epsfig,rotating,setspace,latexsym,amsmath,epsf,amssymb,amsfonts,bm,theorem,cite,enumerate,longtable,accents,float,physics}
\usepackage{algorithm,algorithmic,graphicx,epsf,authblk,epstopdf,url,xcolor, soul,multirow,bbm}
\usepackage{mathtools,comment,tabularx}
\usepackage[center]{qtree}
\usepackage{tree-dvips}
\usepackage[linguistics]{forest }

\newtheorem{theorem}{Theorem}
\newtheorem{corollary}{Corollary}

\newtheorem{remark}{Remark}
\newtheorem{lemma}{Lemma}

\newcommand{\cq}{{\mathcal{Q}}}

\newcommand{\cH}{{\mathcal{H}}}

\newcommand{\ct}{{\mathcal{T}}}

\usepackage{arydshln}

\allowdisplaybreaks

\begin{document}

\title{The Capacity of Semantic Private Information Retrieval with Colluding Servers}
\author{Mohamed Nomeir
\quad  Alptug Aytekin  \quad Sennur Ulukus\\
	\normalsize Department of Electrical and Computer Engineering\\
	\normalsize University of Maryland, College Park, MD 20742 \\
	\normalsize 
    \emph{mnomeir@umd.edu} \quad \emph{aaytekin@umd.edu}   \quad \emph{ulukus@umd.edu}}

\maketitle

\begin{abstract}
  We study the problem of semantic private information retrieval (Sem-PIR) with $T$ colluding servers (Sem-TPIR), i.e., servers that collectively share user queries. In Sem-TPIR, the message sizes are different, and message retrieval probabilities by any user are not uniform. This is a generalization of the classical PIR problem where the message sizes are equal and message retrieval probabilities are identical. The earlier work on Sem-PIR considered the case of no collusions, i.e., the collusion parameter of $T=1$. In this paper, we consider the general problem for arbitrary $T < N$. We find an upper bound on the retrieval rate and design a scheme that achieves this rate, i.e., we derive the exact capacity of Sem-TPIR. 
\end{abstract}

\section{Introduction}
In \cite{chor}, the problem of private information retrieval (PIR) was introduced. In PIR, there are $K$ messages, $W_1, \ldots, W_K$, each of them of the same length $L$ that are replicated among $N$ servers. A user chooses $\theta$ uniformly at random from the set $\{1, \ldots, K\}$, and wishes to retrieve the corresponding message $W_{\theta}$ privately, i.e., without letting any of the servers know the required message index, sends queries to each server. Upon receiving the queries, each server sends an answer based on its dataset and received queries, then transmits it to the user. Upon receiving all answers, the user should be able to decode the required message $W_{\theta}$. In \cite{c_pir}, it was shown that the capacity of this problem, i.e., the highest possible ratio between the number of message bits to the number of downloaded symbols is $C_{PIR}(N,K) = (1+\frac{1}{N}+ \ldots + \frac{1}{N^{K-1}})^{-1}$. In \cite{c_tpir}, another variant of the problem is studied where any $T$ servers can share the queries transmitted by the user, i.e., collude, to decode the required message index. To achieve privacy in the aforementioned problem, $T$-private information retrieval (TPIR), a scheme was designed and the capacity was found as $C_{TPIR}(N,T,K) = C_{PIR}(\frac{N}{T},K)$. There is a rich literature on different variants of the PIR problem \cite{arbitrarycollusion, banawan_eaves, banawan_pir_mdscoded, AG_2, byzantine_tpir, c_spir, ChaoTian, wang_spir, ulukusPIRLC}.

To model a more realistic scenario, \cite{semantic_pir} added two more relaxations to the system model. In \cite{semantic_pir}, the message retrieval probabilities are arbitrary instead of equal priors as in classical PIR or TPIR, and the goal is that the queries do not influence any change in the priors. In addition, the message lengths are arbitrary and can be different for each message. This problem is coined as semantic PIR (Sem-PIR). It was shown that the capacity of Sem-PIR is $C_{Sem-PIR}(N,K) = \frac{\sum_{i=1}^K p_i L_i}{L_1 + \frac{1}{N}L_2 + \ldots + \frac{1}{N^{K-1}}L_K}$, where $p_i$ and $L_i$ are the retrieval probability and the length of the $i$th message, respectively, where without loss of generality, the messages are ordered such that $L_1 \geq L_2 \geq \ldots \geq L_K$.We extend this model and study $T$-colluding Sem-PIR (Sem-TPIR) here. We show that the capacity of Sem-TPIR is $C_{Sem-TPIR}(N,T,K) = C_{Sem-PIR}(\frac{N}{T},K)$ if the priors and message lengths are the same. To show the capacity, we find an upper bound on the achievable rate and provide a scheme that achieves this bound.

\section{Problem Formulation}
Let $N$ denote the number of servers, $K$ the number of messages, and $T < N$ the collusion parameter. Each message symbol is generated uniformly at random, independent for all symbols and for all messages. The length of the $i$th message is $L_i$, and we denote the $i$th message as $W_i$. Thus, the entropy of the messages is given by
\begin{align}
    H(W_1, \ldots, W_K) = \sum_{i=1}^K H(W_i) = \sum_{i=1}^K L_i.
\end{align}
The user sends queries $Q_1^{[\theta]}, \ldots, Q_N^{[\theta]}$ to the $N$ servers to retrieve the $\theta$th message, where $Q_n^{[\theta]}$ denotes the query sent to the $n$th server to retrieve the $\theta$th message. The user has no knowledge of any of the message contents prior to the initiation of the scheme, thus,
\begin{align}
    I(W_1,\ldots,W_K ; \cq) = 0,
\end{align}
where $\cq = \{Q_n^{[\theta]}, ~n \in [N], ~\theta \in [K]\}$.

The user does not know which $T$ of the $N$ servers are colluding. This implies that we need to make sure that any $\ct \in \{1, \ldots, N\}$, with $|\ct| = T$ servers do not know the required message index from the transmitted queries, thus, 
\begin{align}\label{privacy_const_1}
    I(\Theta; Q^{[\theta]}_{\ct}) = 0, ~ \theta \in [K],
\end{align}
or equivalently 
\begin{align}\label{privacy_const_2}
    \mathbb{P}(\Theta = \theta | Q^{[\theta]}_{\ct}) = \mathbb{P}(\Theta = \theta) = p_{\theta}, ~ \theta \in [K]. 
\end{align}
Upon receiving the queries, the honest but curious servers compute their individual answers based on the messages and the received queries, thus,
\begin{align}
    H(A_n^{[\theta]}|Q_n^{[\theta]}, W_1, \ldots, W_K) = 0,  ~ \theta \in [K],\label{deterministicanswer}
\end{align}
where $A_n^{[\theta]}$ is the answer computed by server $n$ for query $Q_n^{[\theta]}$. Finally, upon receiving the answers from all servers, the user must decode the required message index, thus,
\begin{align}\label{decodability}
    H(W_{\theta}| A_{[1:N]}^{[\theta]},  Q_{[1:N]}^{[\theta]})=0,  ~ \theta \in [K].
\end{align}
The rate of Sem-TPIR is defined as the ratio between the average message length and the average number of downloaded symbols, 
\begin{align}
    R_{Sem-TPIR}(N,T,K, \{L_i\}_{i \in [K]}, \{p_{i}\}_{i=1}^K) = \frac{\mathbb{E}[L]}{\mathbb{E}[D]},
\end{align}
where the expected value is over the message retrieval distribution $p_1, \ldots, p_K$. The capacity is defined as the highest possible achievable rate over all possible retrieval schemes $\Pi$ that satisfy \eqref{privacy_const_1}-\eqref{decodability}, that is, 
\begin{align}
    &C_{Sem-TPIR}(N,T,K, \{L_i\}_{i \in [K]}, \{p_{i}\}_{i=1}^K) \nonumber \\ 
    & \quad =\sup_{\Pi} R_{Sem-TPIR}(N,T,K, \{L_i\}_{i \in [K]}, \{p_{i}\}_{i=1}^K).
\end{align}
\begin{remark}
    Note that \eqref{privacy_const_1}, \eqref{privacy_const_2} and \eqref{deterministicanswer} imply 
    \begin{align}
        I(\Theta; Q_{\ct}^{[\theta]},A_{\ct}^{[\theta]}| W_{1}, \ldots, W_K) &= 0, ~\quad \theta \in [K]\\
        \mathbb{P}(\Theta = \theta| Q_{\ct}^{[\theta]}, A_{\ct}^{[\theta]}, W_{1}, \ldots, W_K ) &= \mathbb{P}(\Theta = \theta) = p_{\theta}.
    \end{align}
\end{remark}

\begin{remark}\label{remark_queries_answers}
    Since $Q_{\ct}^{[\theta]}$ does not convey any information about the required message index, $A_{\ct}^{[\theta]}$ must be independent of the message index for any private retrieval scheme, i.e.,
    \begin{align}
        H(A_{\ct}^{[1]}| \cq) =  \ldots = H(A_{\ct}^{[K]}| \cq) = H(A_{\ct}| \cq). 
    \end{align}
\end{remark}

\section{Main Result}
\begin{theorem}\label{main_thm}
   Let $N$ be the number of servers and $K$ be the number of messages that are replicated among the servers. Let $T$ be the collusion parameter. Then, the capacity of the private information retrieval is given by
   \begin{align}
       &C_{Sem-TPIR}(N,T,K, \{L_i\}_{i \in [K]}, \{p_{i}\}_{i=1}^K) \nonumber \\ 
       & \quad =  \frac{\mathbb{E}[L]}{L_1 + (\frac{T}{N})L_2 + \ldots + (\frac{T}{N})^{K-1}L_K},
   \end{align}
   where $L_i$ are the lengths of the messages, $p_i$ are the retrieval priors, where $L_1 \geq L_2 \geq \ldots \geq L_K$.
\end{theorem}

\section{Corollaries and Discussions}
In this section, important connections between semantic TPIR and previous variants in the literature are considered. As evident, the foremost candidate for comparison is the TPIR with equal message lengths and equal priors. In that regard, we have the following intuitively pleasing corollaries. 

\begin{corollary}
    The capacity of semantic TPIR is higher than the capacity of TPIR with equal message sizes when the following condition is satisfied
    \begin{align}
        \sum_{i=1}^K (L_i - \mathbb{E}[L]) \left(\frac{T}{N}\right)^{i-1} \leq 0.
    \end{align}
\end{corollary}

\begin{corollary}
    The capacity of Sem-TPIR is always higher than the rate of TPIR with zero padding.
\end{corollary}

The previous two corollaries are extensions of the corollaries in \cite{semantic_pir} to the case of $T$-colluding. However, in an unusual manifestation, the capacity of Sem-TPIR can be higher than the capacity of PIR of fixed message sizes and PIR with zero padding as shown in the following two corollaries and examples.

\begin{corollary}
    The capacity of Sem-TPIR is higher than the capacity of PIR when the following is satisfied
    \begin{align}
        \sum_{i=1}^K (\mathbb{E}[L] - T^{i-1}L_i)\frac{1}{N^{i-1}} \geq 0.
    \end{align}
\end{corollary}

As a simple numerical example, for the case of $N=10$ servers and $K=2$ messages, the classical PIR capacity is equal to $0.9081$. However, for the same case with $T=2$, $L_1 = 1000$, $L_2 = 100$, with probabilities $0.99$ and $0.01$, the Sem-TPIR capacity is $0.9716$.

\begin{corollary}
    The capacity of Sem-TPIR is higher than the zero padding rate of the PIR when the following conditions are satisfied
    \begin{align}
        L_1 > T^{i-1} L_i, \quad i \in \{2, \ldots, K\}.
    \end{align}
\end{corollary}

\section{Achievable Scheme}
Let $N$ be the number of servers, $T$ be the collusion parameter and $K$ be the number of messages, with $\theta$ being the required message index. Let the messages be ordered in decreasing order based on their length, i.e., $L_1 \geq \ldots \geq L_K$. First, we sub-packetize each message into $U_1, \ldots, U_K$ symbols where $L_i = \alpha U_i$. The scheme steps are as follows for each sub-packetization. First, let the symbols of each message downloaded at each iteration of the scheme be denoted as $W_i$, then:
\begin{itemize}
    \item First choose $S_1, S_2, \ldots, S_K$ square invertible matrices uniformly at random of size $U_1, U_2, \ldots, U_K$, respectively. Let the new message symbols be $W_i' = S_i W_i$.
    \item Step 1 (Singletons): Download $N\nu_i$, $i \in [K]$, symbols for each message from the $N$ servers in the following way. If $i = \theta$, download $W'_i(1:N\nu_i)$ from the $N$ servers equally, i.e., $\nu_i$ from each server. If $i \neq \theta$, apply $MDS_{N(\nu_i + \frac{N-T}{T}\min(\nu_i,\nu_{\theta})),N\nu_i}$ on $W'_i$ and download the first $N\nu_i$ symbols equally as well. 

    \item Step 2 ($s$-Sum): For each $s$, where $2 \leq s \leq K$, download $ (\frac{N-T}{T})^{s-1} \min(\nu_{\mathcal{S}})$ $s$-linear combinations of the message symbols in $\mathcal{S}$, for all $|\mathcal{S}| = s$, where $\nu_{\mathcal{S}}=\{\nu_{p}\}_{p\in\mathcal{S}}$, such that the following are satisfied:
    \begin{enumerate}
        \item If $\theta \notin \mathcal{S}$, let $\nu_{i_k}=\min(\nu_\mathcal{S})$. Code the fresh symbols of $W'_s,\ s \in \mathcal{S}$ according to $MDS_{N(\frac{N-T}{T})^{s-2}(v_{i_k}+\frac{N-T}{T}\min\{\nu_{i_k},\nu_{\theta}\})\times N(\frac{N-T}{T})^{s-2}v_{i_k}}$. Then, download $N(\frac{N-T}{T})^{s-2}v_{i_k}$ symbols from the sum of these fresh symbols.
        
        \item If $\theta \in \mathcal{S}$, two cases emerge. 
        \begin{itemize}
            \item Case I: $\theta \neq \textrm{argmin}(\nu_{\mathcal{S}})$ 
            
                Let $\mathcal{S} = \{\theta, i_1, \ldots, i_{s-1} \} $ with $i_k$ being the index such that $\nu_{i_k} = \min(v_{\mathcal{S}})$. Download new symbols of $W'_{\theta}$ using its sums with $N\nu_{i_k}(\frac{N-T}{T})^{s-2} \frac{N}{T}$ coded symbols remaining from previous step, i.e., $(s-1)$st step.

            \item Case II: $\theta = \textrm{argmin}(v_{\mathcal{S}})$

            Let $\mathcal{S}\setminus\theta = \{i_1, \ldots, i_{s-1}\}$  with $i_k$ being the index such that $\nu_{i_k} = \min(v_{\mathcal{S}\setminus \theta})$. Download new symbols of $W'_{\theta}$ using its sums with $N\nu_{\theta}(\frac{N-T}{T})^{s-2} \frac{N}{T}$ coded symbols remaining from previous step.
        \end{itemize}
    \end{enumerate}
    \item Step 3: Repeat this procedure $\alpha$ times ($\alpha$ is specified later).
\end{itemize}
Thus, the number of downloaded symbols each time is
\begin{align}
    D =&  N  \sum_{i=1}^K \nu_i + N \sum_{s=2} ^K \sum_{i=s}^K{i-1 \choose s-1}\left( \frac{N-T}{T}\right)^{s-1} \nu_i\\
    =& N  \sum_{i=1}^K \nu_i + N \sum_{i=2} ^K \sum_{s=2}^i{i-1 \choose s-1}\left( \frac{N-T}{T}\right)^{s-1} \nu_i\\
    =& N  \sum_{i=1}^K \nu_i + N \sum_{i=2} ^K \nu_i \left(\sum_{s=0}^{i-1} \left(\frac{N-T}{T}\right)^s {i-1 \choose s} -1\right)\\
    =& N  \sum_{i=1}^K \nu_i + N \sum_{i=2}^K \nu_i \left(\frac{N^{i-1}}{T^{i-1}} - 1 \right)\\
    =& \sum_{i=1}^K \frac{N^i}{T^{i-1}}\nu_i,
\end{align}
and the number of symbols for the required message index is given by
\begin{align}
    U_{\theta} =& N \nu_{\theta} + \sum_{s=2}^{\theta} N (\frac{N-T}{T})^{s-1} \nu_{\theta} {\theta -1 \choose s-1} \nonumber \\  &+ \sum_{s=2}^{\theta} \sum_{i=\theta + 1}^{K} N (\frac{N-T}{T})^{s-1} \nu_{i}  {i -2 \choose s-2} \nonumber\\
    & + \sum_{s= \theta +1}^K \sum_{i=s}^{K} N (\frac{N-T}{T})^{s-1} \nu_i {i -2 \choose s-2}\\
    =& N \nu_{\theta} + N \nu_{\theta} \left(\sum_{i=0}^{\theta -1 } (\frac{N-T}{T})^{i}{\theta -1 \choose i} -1\right) \nonumber \\ &+ N\left( \sum_{i=\theta+1}^K \frac{N-T}{T}\nu_i {i-2 \choose 0} \nonumber \right.\\
    & \left. + \sum_{i=\theta+1}^K (\frac{N-T}{T})^{2}\nu_i {i-2 \choose 1}  \nonumber \right. \\  & \left.  + \ldots + \sum_{i=\theta+1}^K (\frac{N-T}{T})^{\theta -1 }\nu_i {i-2 \choose \theta -2}    \right) \nonumber\\
    &+ N \left( \sum_{i = \theta +1}^{K} (\frac{N-T}{T})^{\theta} \nu_i {i-2 \choose \theta -1} \nonumber \right. \\ 
    & \left. + \sum_{i = \theta +2}^K (\frac{N-T}{T})^{\theta + 1} \nu_i {i-2 \choose \theta } \nonumber \right. \\  & \left. + \ldots + (\frac{N-T}{T})^{K-1} \nu_K {K-2 \choose K-2} \right) \\
    =& \frac{N^{\theta}}{T^{\theta -1 }} \nu_{\theta} + \frac{N}{T}(N-T) \left( \nu_{\theta +1} \sum_{i=0}^{\theta -1} (\frac{N-T}{T})^i{\theta -1 \choose i}\nonumber \right.\\ & \left.+ \nu_{\theta + 2} \sum_{i=0}^{\theta} (\frac{N-T}{T})^i{\theta \choose i}+ \ldots \nonumber \right. \\  & \left. + \nu_K \sum_{i=0}^{K-2} (\frac{N-T}{T})^i{K-2 \choose i} \right)\\ 
    =& \frac{N^{\theta}}{T^{\theta -1 }} \nu_{\theta} + (N-T)\sum_{i=\theta+1}^{K} \left(\frac{N}{T}\right)^{i-1}\nu_i.
\end{align}

Thus, the relation between message symbols $U_i$, and $\nu_i$, where $i \in [K]$, is given by the following 
\begin{align}
    &\begin{bmatrix}
        U_1,
        \ldots,
        U_K
    \end{bmatrix}^t = V 
\begin{bmatrix}
    \nu_1,
    \ldots,
    \nu_K
\end{bmatrix}^t,
\end{align}
where
\begin{align}
    V = \begin{bmatrix}
        N& (N-T)\frac{N}{T}& (N-T)\frac{N^2}{T^2}& \ldots& (N-T)\frac{N^{K-1}}{T^{K-1}}\\
0& \frac{N^2}{T}& 
(N-T)\frac{N^2}{T^2}& \ldots& (N-T)\frac{N^{K-1}}{T^{K-1}}\\
\vdots& \vdots& \vdots& \ddots& \vdots\\
0& 0& 0& \ldots& \frac{N^K}{T^{K-1}}\end{bmatrix}
\end{align}

From this, $\nu_1, \ldots, \nu_K$, can be computed as follows
\begin{align}\label{compute_nus}
    \begin{bmatrix}
        \nu_1,
        \ldots,
        \nu_K
    \end{bmatrix}^t = \frac{1}{\alpha} \underbrace{ V^{-1}
\begin{bmatrix}
    L_1,
    \ldots,
    L_K
\end{bmatrix}^t}_{M},
\end{align}
where $\alpha = \gcd(L_{[K]},M([K]))$, and
\begin{align}
    V^{-1} = \begin{bmatrix}
        \frac{1}{N}& \frac{-(N-T)}{N^2}& \frac{-(N-T)T}{N^3}& \ldots& \frac{-(N-T)T^{K-2}}{N^K}\\
0& \frac{T}{N^2}& 
\frac{-(N-T)T}{N^3}& \ldots& \frac{-(N-T)T^{K-2}}{N^K}\\
\vdots& \vdots& \vdots& \ddots& \vdots\\
0& 0& 0& \ldots& \frac{T^{K-1}}{N^K}\end{bmatrix}
\end{align}

Now, the expected value of the downloaded symbols for our scheme is given by
\begin{align}
    \alpha \mathbb{E}[D] =& \alpha D = \alpha \sum_{i=1}^K \frac{N^i}{T^{i-1}}\nu_i\\
    =& \sum_{i=1}^K \frac{N^i}{T^{i-1}} \left( \frac{T^{i-1}}{N^i}L_i - (N-T) \sum_{j=i+1}^K \frac{T^{j-2}}{N^j} L_j \right)\\
    =& \sum_{i=1}^K L_i - (N-T) \sum_{i=2}^K \left(\frac{1}{N}+ \frac{T}{N^2}+ \ldots+ \frac{T^{i-2}}{N^{i-1}}\right)L_i\\
    =& \sum_{i=1}^K \frac{T^{i-1}}{N^{i-1}}L_i
\end{align}

and the rate is given by 
\begin{align}
    R &= \frac{\alpha \sum_{i=1}^K p_i U_i}{\alpha D}= \frac{\mathbb{E}[L]}{L_1 + \frac{T}{N}L_2 + \ldots + \frac{T^{K-1}}{N^{K-1}}L_K} \nonumber \\ &= C_{Sem-TPIR}(N,T,K,\{L_i\}_{i \in [K]}, \{p_{i}\}_{i=1}^K).
\end{align}

\begin{remark}
    To make sure that for the $s$-sum, $(\frac{N-T}{T})^{s-1}\nu_k$ are always positive integers we first note that $k \geq s$ for the $s$-sum. Then, we proceed as follows
    \begin{align}
       & \left(\frac{N-T}{T}\right )^{s-1}\nu_s \nonumber \\=&  \frac{1}{\alpha}\left(\frac{N-T}{T}\right)^{s-1} \left(\frac{T^{s-1}}{N^{s}}L_s - \sum_{j=s+1}^K \frac{(N-T)T^{j-2}}{N^j}L_j\right) \\
        =& \frac{1}{\alpha} (N-T)^{s-1} \Big(\beta_s N^{K-s} -   \sum_{j=s+1}^K \beta_j N^{K-j} (N-T)T^{j-s-1}\Big) \label{remerk_1_comment}
    \end{align}
    where $L_i = \beta_i N^K$. Now, since $ 1 \leq s \leq K$, and $s+1 \leq j \leq K$, we guarantee that $(N-T)^{s-1} \left(\beta_s N^{K-s} -  \sum_{j=s+1}^K \beta_j N^{K-j} (N-T)T^{j-s-1}\right)$ is an integer (still not proven to be positive). To prove it is a positive integer, recall that $\beta_j \leq \beta_s, ~ j \geq s+1$, thus 
    \begin{align}
        \sum_{j=s+1}^K & \beta_j N^{K-j} (N-T)T^{j-s-1} \nonumber \\  & \leq   \beta_s (N-T) \frac{N^K}{T^{s+1}} \sum_{j=s+1}^K  (\frac{T}{N})^j \\
        & =  \beta_s (N-T) \frac{N^K}{T^{s+1}} \sum_{j=0}^{K-s-1} (\frac{T}{N})^{j+s+1}\\
        &=  \beta_s (N-T) N^{K-s-1} \sum_{j=0}^{K-s-1} (\frac{T}{N})^{j}\\
        & \leq \beta_s (N-T) N^{K-s-1} \sum_{j=0}^{\infty} (\frac{T}{N})^{j} \\
        & = \beta_s (N-T) N^{K-s-1} \frac{N}{N-T}\\
        & = \beta_s N^{K-s},
    \end{align}
    and therefore, \eqref{remerk_1_comment} is a positive integer.
\end{remark}

\begin{remark}
    Note that, in our scheme, to make sure that the $s$-sum of the new interference symbols are compatible, we use the same MDS code. This will be more evident with the illustrative examples provided next.
\end{remark}

\section{Illustrative Examples}
\subsection{Example 1}
Let $N=4$ servers, $T=3$ colluding parameter, $K=3$ messages, with $L_1 = 192$, $L_2 = 128$, and $L_3= 64$ symbols, and probabilities $p_1=\frac{1}{2}, ~ p_2 = \frac{1}{3}, ~ p_3 = \frac{1}{6}$. Thus, based on the scheme presented in the previous section, we have the following parameters: $\alpha = 1$, $\nu_1= 37$, $\nu_2 = 21$, $\nu_3 = 9$. Finally, let $S_i$, $i \in [3]$ be square invertible matrices of sizes, 192, 128, and 64 chosen uniformly at random with denoting $W'_i = S_iW_i$. The retrieval schemes for $W_1$, $W_2$, and $W_3$ are given in Table \ref{example_1_m1}. When retrieving $W_1$, we have 
\begin{align}
    a_{[1:192]} =& W'_1 = S_1 W_1\\
    b_{[1:112]} =& MDS_{112\times 84} W'_2(1:84)\\
    b_{[113:128]} =& MDS_{16 \times 12}  W'_2(85:96)\\
    c_{[1:48]} =& MDS_{48 \times 36}W'_3(1:36)\\
    c_{[49:64]}=& MDS_{16 \times 12} W'_3(37:48).
\end{align}

When retrieving $W_2$, we have
\begin{align}
    a_{[1:176]} =& MDS_{176 \times 148} W'_1 (1:148)\\
    a_{[177:192]} =& MDS_{16 \times 12}W'_1 (149:160)\\
    b_{[1:128]} =&W'_2 =  S_2 W_2\\
    c_{[1:36],[49:60]} =& MDS_{48 \times 36}W'_3(1:36)\\
    c_{[37:48],[61:64]}=& MDS_{16 \times 12} W'_3(37:48).
\end{align}

Finally, when retrieving $W_3$, we have
\begin{align}
    a_{[1:148],[177:188]} =& MDS_{160 \times 148} W'_1(1:148)\\
    a_{[149:176],[189:192]} =& MDS_{32 \times 28} W'_1 (149:160)\\
    b_{[1:84],[113:124]} =& MDS_{112 \times 84} W'_2(1:84)\\
    b_{[85:112],[125:128]} =& MDS_{32 \times 28}W'_2(85:96)\\
    c_{[1:64]}=& W'_3 = S_3 W_3.
\end{align}

\begin{table}[h]
    \centering
    \begin{tabular}{|c|c|c|c|}
         \hline
         DB 1& DB 2& DB 3& DB 4\\
         \hline
         $\begin{aligned}
             a_{[1:37]}
         \end{aligned}$& $a_{[38:74]}$& $a_{[75:111]}$& $a_{[112:148]}$\\
         $b_{[1:21]}$& $b_{[22:42]}$& $b_{[43:63]}$& $b_{[64:84]}$\\
         $c_{[1:9]}$ & $c_{[10:18]}$ & $c_{[19:27]}$ & $c_{[28:36]}$\\
         \hline
        $\begin{aligned}
            a_{[149:155]}\\+b_{[85:91]}
        \end{aligned}$ &$\begin{aligned}a_{[156:162]}\\+b_{[92:98]}\end{aligned}$&$\begin{aligned}a_{[163:169]}\\+b_{[99:105]}\end{aligned}$&$\begin{aligned}a_{[170:176]}\\+b_{[106:112]}\end{aligned}$\\
         \hdashline
         $\begin{aligned}a_{[177:179]}\\+c_{[37:39]}\end{aligned}$&$\begin{aligned}a_{[180:182]}\\+c_{[40:42]}\end{aligned}$&$\begin{aligned}a_{[183:185]}\\+c_{[43:45]}\end{aligned}$&$\begin{aligned}a_{[186:188]}\\+c_{[46:48]}\end{aligned}$\\
\hdashline
         $\begin{aligned}b_{[113:115]}\\+c_{[49:51]}\end{aligned}$ & $\begin{aligned}b_{[116:118]}\\+c_{[52:54]}\end{aligned}$& $\begin{aligned}b_{[119:121]}\\+c_{[55:57]}\end{aligned}$ & $\begin{aligned}b_{[122:124]} \\+c_{[58:60]}\end{aligned} $ \\
         \hline
         $\begin{aligned}
             a&_{189} +b_{125}\\&+c_{61}
         \end{aligned}$ & $\begin{aligned} a&_{190}+b_{126}\\&+c_{62}\end{aligned}$ & $\begin{aligned}a&_{191}+b_{127}\\&+c_{63}\end{aligned}$ & $\begin{aligned}a&_{192}+b_{128}\\ &+c_{64}\end{aligned}$ \\  
         \hline
    \end{tabular}
    \vspace*{0.2cm}
    \caption{Retrieval Scheme for Example 1.}
    \label{example_1_m1}
\end{table}

The average rate of the scheme is then given by
\begin{align}
    R =& \frac{1}{2} R_1 + \frac{1}{3}R_2 + \frac{1}{6}R_3\\
    =& \frac{\frac{1}{2}L_1+\frac{1}{3}L_2+\frac{1}{6}L_3}{324} =  \frac{\mathbb{E}[L]}{324}.
\end{align}
The optimal rate is given by
\begin{align}
    C_{Sem-TPIR}(4,3,3) =& \frac{\mathbb{E}[L]}{192+(\frac{3}{4})128 + (\frac{16}{9})64}\\
    =& \frac{\mathbb{E}[L]}{324}.
\end{align}
Thus, the average rate of the developed scheme achieves the capacity.

\subsection{Example 2 }
Let $N=8$ servers, $T=2$ colluding parameter, $K=4$ messages with lengths $L_1 = 16384$, $L_2 = 12288$, $L_3 = 8192$, and $L_4 = 4096$. Using the scheme developed, we have the following retrieval parameters: $\alpha=8$, $U_1 = 2048$, $U_2 = 1536$, $U_3 = 1024$, $U_4 = 512$, with $\nu_1 = 85$, $\nu_2 = 21$, $\nu_3 = 5$, and $\nu_4 = 1$. To make it easier to visualize the retrieval scheme in Table \ref{example_2_m1}, we put the numbers of downloaded symbols and the combinations of the messages related to these numbers. 

\begin{table}[h]
    \centering
    \begin{tabular}{|c|c|c|c|c|c|c|c|c|}
         \hline
         Combinations &DB 1& DB 2& $\ldots$ & DB  8\\
         \hline
         $W_1$& 85&85& $\ldots$ &85\\
         \hdashline
         
$W_2$& 21& 21& $\ldots$ &21\\
\hdashline

$W_3$ & 5& 5&$\ldots$&5\\
\hdashline

$W_4$& 1&1&$\ldots$&1\\
\hline
$W_1 \sim W_2$& 63& 63&$\ldots$&63\\
\hdashline
$W_1 \sim W_3$&15&15&$\ldots$&15\\
\hdashline
$W_1\sim W_4$&3&3&$\ldots$&3\\
\hdashline
$W_2 \sim W_3$& 15& 15& $\ldots$ & 15\\
\hdashline
$W_2 \sim W_4$ & 3& 3& $\ldots$& 3\\
\hdashline
$W_3 \sim W_4$ & 3& 3& $\ldots$& 3\\
\hline
$W_1 \sim W_2 \sim W_3$& 45& 45& $\ldots$& 45\\
\hdashline
$W_1 \sim W_2 \sim W_3$ & 9 & 9&$\ldots$ & 9\\
\hdashline
$W_2 \sim W_3 \sim W_4$ & 9 & 9&$\ldots$ & 9\\
\hdashline
$W_1 \sim W_3 \sim W_4$ & 9 & 9&$\ldots$ & 9\\
\hline
$W_1 \sim W_2 \sim W_3 \sim W_4$ & 27 & 27& $\ldots$& 27\\
\hline
    \end{tabular}
    \vspace*{0.2cm}
    \caption{Retrieval Scheme for Example 2.}
    \label{example_2_m1}
\end{table}

The rate achieved using our scheme is $R = \frac{\mathbb{E}[L]}{8 \times 2504}$, and the capacity is $C_{Sem-TPIR} = \frac{\mathbb{E}[L]}{8 \times 2504} = R$.

\section{Privacy Proof}
First note that for $MDS_{a \times b}$, any $b$ columns are independent. In addition, \cite[Lemma~1]{c_tpir} with some dimension manipulations yields the following corollary. 
\begin{corollary}
    Let $S_i \in GL_q(U_i)$, $i \in [K]$, chosen uniformly at random and $G_i \in GL_q(\beta_i)$, $i \in [K]$ with $\mathcal{I}_i \subset [U_i]$, with $|\mathcal{I}_i| = \beta_i$. Then, the following two distributions are equivalent
    \begin{align}
        &\left(G_1S_1(\mathcal{I}_1,:), \ldots, G_KS_K(\mathcal{I}_K,:) \right) \nonumber \\ 
        &\quad\sim \left(S_1([\beta_1],:]), \ldots, S_1([\beta_K],:]) \right). 
    \end{align}
\end{corollary}
This shows that the servers will not recognize the difference between the MDS-coded interference and the pure symbols required to retrieve the message index.

Finally, in our scheme, we use fresh interference symbols from the $(s-1)$-sum phase to decode new symbols for the required message index in the $s$-sum phase. Thus, we need to make sure that any $T$ servers that share the fresh interference symbols from the $(s-1)$-sum along with the symbols used in the $s$-sum step appear independent from each other. To prove this, let $\mathcal{S}' = \{i_1, \ldots, i_{s-1}\}$ be the indices of the messages that new interference symbols are downloaded in the $(s-1)$-sum step, with $\nu_{i_k} = \min(\nu_{\mathcal{S}'})$. Thus, the number of fresh interference symbols shared among the colluding servers is $T (\frac{N-T}{T})^{s-2} \nu_{i_k}$. Let $\mathcal{S} = \{\theta\} \cup \mathcal{S}'$, we have two different cases. The first case is $\nu_{i_k} = \min(\nu_{\mathcal{S}})$, thus the number of the shared downloaded symbols, among the colluding servers, used in interference for the $s$-sum phase is $T (\frac{N-T}{T})^{s-1} \nu_{i_k}$. Now, the total number of shared symbols is $T (\frac{N-T}{T})^{s-2} \nu_{i_k} (\frac{N}{T}) = N (\frac{N-T}{T})^{s-2} \nu_{i_k}$, which is equal to the number of independent columns in the $MDS$ encoding used in our scheme, thus they appear independent for any $T$ servers. In the second case, we have $\nu_{\theta} = \min(\nu_{\mathcal{S}})$, thus the number of shares symbols in the $s$-sum phase related the $(s-1)$-sum phase is $T(\frac{N-T}{T})^{s-1}\nu_{\theta}$. Thus, the total number of shared symbols is $T(\frac{N-T}{T})^{s-2} (\nu_{i_k}+ (\frac{N-T}{T})\nu_{\theta}) \leq T(\frac{N-T}{T})^{s-2} \nu_{i_k}(\frac{N}{T}) = N (\frac{N-T}{T})^{s-2} \nu_{i_k}$. Thus, in this case as well the number of downloaded symbols collectively is less than the number of columns of the $MDS$ used in encoding, which ensures privacy. This shows that the scheme used appear symmetric for any $T$ colluding servers for any $\theta \in [K]$ ensuring privacy. 

\section{Converse Proof}
We start with the definitions:
\begin{align}
    \cq =& \{Q_{n}^{[\theta]}, ~n \in [N], ~\theta \in [K]\},\\
    A_{\mathcal{T}}^{[\theta]} =& \{A_n^{[\theta]}, ~n \in \mathcal{T}\},\\
    \cH_T =& \frac{1}{{N \choose T}}\sum_{\mathcal{T} \subset [N]: |\mathcal{T}| = T} \frac{H(A_{\mathcal{T}}| \cq)}{T}.
\end{align}

For completeness, we restate Han's inequality \cite{coverthomas},

\begin{lemma}[Han's Inequality]
    \begin{align}
        \cH_T \geq \frac{H(A_{1}^{[\theta]}, \ldots, A_{N}^{[\theta]}| \cq)}{N}.
    \end{align}
\end{lemma}

In addition, we provide the following result.

\begin{lemma}
    \begin{align}
        N\cH_T \leq \sum_{n \in [N]} H(A_n | \cq).
    \end{align}
\end{lemma}

To start with the converse proof, first consider the simple case with $K=1$ or $K=2$ messages.

\paragraph{Case 1: $K=1$ and arbitrary $N$}
Let $L_i$ be the length of the message $W_i \in \{W_1, \ldots, W_K\}$, then
\begin{align}
    L_i &= H(W_i) = H(W_i|\cq)
     = I(W_i; A_{1}, \ldots,A_{N}|\cq)\\
    & = H(A_{1}, \ldots,A_{N}|\cq) \leq N\cH_T.
\end{align}
\paragraph{Case 2: $K=2$ and arbitrary $N$}
Let $L_i, L_j$ be the lengths of the messages $W_i, ~W_j \in \{W_1, \ldots, W_K\}$, with $i \neq j$. Then, using the same steps as \cite{c_tpir}, we have
\begin{align}
    L_i+L_j &= H(W_i,W_j) = H(W_i,W_j|\cq) \\
& \leq  N\cH_T + L_j - H(A_{\ct}| W_i,\cq) \label{c_1}.
\end{align}
By averaging over all possible $\ct$, we have
\begin{align}
    L_i &\leq N\cH_T - \sum_{\ct \subset [N]: |\ct| = T} H(A_{\ct}| W_i,\cq)\\
    &\leq  N\cH_T - \frac{T}{N} H(A_{[N]}^{[j]} | W_1, \cq)\\ 
    &= N\cH_T - \frac{T}{N} L_j.
\end{align}
Thus, since the proof is symmetric over $i$ and $j$, we have
\begin{align}
    N\cH_T \geq \max\left(L_i + \frac{T}{N} L_j, L_j + \frac{T}{N} L_i\right)
\end{align} 
\paragraph{Case 3: Arbitrary $K$ and arbitrary $N$}
We proceed similarly to the previous two cases as follows. First, choose any arbitrary permutation $(i_1,\ldots, i_K)$ of $[K]$, then
\begin{align}
    \sum_{j=1}^K L_{i_j}  =& H(W_{i_1},W_{i_2}, \ldots , W_{i_K}|\cq)\\
     =& I(A_{\ct}, A_{\overline{\ct}}^{[1]}, \ldots, A_{\overline{\ct}}^{[K]}; W_{i_1},W_{i_2}, \ldots , W_{i_K}| \cq )\\
    \leq & N \cH_T + \sum_{n \in \overline{\ct}}H(A_n^{[i_2]}|A_{\ct}, W_{i_1}, \cq) + \sum_{j=3}^K L_{i_j}\nonumber \\ & - H(A_{\ct}| W_{i_1}, W_{i_2},\cq).
\end{align}

Now, we have
\begin{align}
&L_{i_1}+L_{i_2} + H(A_{\ct}| W_{i_1}, W_{i_2},\cq) \nonumber \\ & \quad \leq N \cH_T + \sum_{n \in \overline{\ct}}H(A_n^{[i_2]}|A_{\ct}, W_{i_1}, \cq)
\end{align}

By averaging over all possible subsets $\ct$, we have
\begin{align}
    &L_{i_1}+L_{i_2} + \frac{1}{{N \choose T}} \sum_{\ct}H(A_{\ct}| W_{i_1}, W_{i_2},\cq) \nonumber \\ & \leq  N \cH_T + \frac{1}{{N \choose T}} \sum_{\ct}\sum_{n \in \overline{\ct}}H(A_n^{[i_2]}|A_{\ct}, W_{i_1}, \cq)\\
    & \leq  N \cH_T + \left(\frac{N}{T}-1\right)\left(N\cH_T - L_{i_1}\right).
\end{align}

Upon rearranging, we have
\begin{align}
    N\cH_T \geq& L_{i_1} + \frac{T}{N}L_{i_2} + \frac{1}{{N \choose T}} \frac{NT^2}{N^2}\sum_{\ct}\frac{H(A_{\ct}| W_{i_1}, W_{i_2},\cq)}{T}\\
    \geq& \ldots  \geq  L_{i_1} + \frac{T}{N}L_{i_2}+ \frac{T^2}{N^2}L_{i_3}+ \ldots + \frac{T^{K-1}}{N^{K-1}}L_{i_K}.
\end{align}
Since $(i_1, i_2, \ldots, i_K)$ is an arbitrary permutation for $[K]$, then,
\begin{align}
    N\cH_T \geq \max_{\mathcal{P}_K} \left( L_{i_1} + \frac{T}{N}L_{i_2}+ \frac{T^2}{N^2}L_{i_3}+ \ldots + \frac{T^{K-1}}{N^{K-1}}L_{i_K}\right),
\end{align}
which is maximum when $L_{i_1} \geq L_{i_2} \geq \ldots \geq L_{i_K}$. Thus,
\begin{align}
    R =& \frac{\mathbb{E}[L]}{\mathbb{E}[D]} \\
    =& \frac{\mathbb{E}[L]}{\sum_{i=1}^K p_i \sum_{n=1}^N H(A_n^{[i]})}\\
    =&\frac{\mathbb{E}[L]}{ \sum_{n=1}^N H(A_n^{[i]})} \\
    \leq &\frac{\mathbb{E}[L]}{ \sum_{n=1}^N H(A_n^{[i]}|\cq)}\\
    \leq &\frac{\mathbb{E}[L]}{ N\cH_T}\\
    \leq & \frac{\mathbb{E}[L]}{\left( L_{1} + \frac{T}{N}L_{2}+ \frac{T^2}{N^2}L_{3}+ \ldots + \frac{T^{K-1}}{N^{K-1}}L_{K}\right)},
\end{align}
with $L_1 \geq \ldots \geq L_K$.

\bibliographystyle{unsrt}
\bibliography{references}
\end{document}